\documentclass[preprint,12pt]{elsarticle}

\usepackage[breaklinks=true]{hyperref}
\usepackage{breakcites}
\usepackage{natbib}
\usepackage{amssymb}

\biboptions{authoryear}

\newcommand{\pd}{p_{d}}
\newcommand{\pr}{p_{r}}

\newcommand{\dt}{\Delta t}
\newcommand{\wo}{w_0}
\newcommand{\ptri}{p_\Delta}

\begin{document}

\begin{frontmatter}

\title{Social Physics: Uncovering Human Behaviour from Communication}

\author{Kunal Bhattacharya\fnref{fn1}}

\author{Kimmo Kaski\fnref{fn2}}

\address{Department of Computer Science, Aalto University School of Science, P.O.Box 15500, FI-00076 AALTO, Finland}

\fntext[fn1]{\href{kunal.bhattacharya@aalto.fi}{\nolinkurl{kunal.bhattacharya@aalto.fi}}}
\fntext[fn2]{\href{kimmo.kaski@aalto.fi}{\nolinkurl{kimmo.kaski@aalto.fi}}}




\begin{abstract}
In the post year 2000 era the technologies that facilitate human communication have rapidly multiplied. While the adoption of these technologies has hugely impacted the behaviour and sociality of people, specifically in urban but also in rural environments, their ``digital footprints''  on different data bases have become an active area of research. The existence and accessibility of such large population-level datasets, has allowed scientists to study and model innate human tendencies and social patterns in an unprecedented way that complements traditional research approaches like questionnaire studies. In this review we focus on data analytics and modelling research - we call Social Physics - as it has been carried out using the mobile phone data sets to get insight into the various aspects of human sociality, burstiness in communication, mobility patterns, and daily rhythms. 
\end{abstract}


\end{frontmatter}

\clearpage

\section{Introduction}
Technologies that harness digital information have grown over the past two decades in a number of unprecedented ways. It is said that in the current state-of-the-art, everyday an order of $3 \times 10^{18}$ bytes of data gets created\footnote{\url{https://www-01.ibm.com/common/ssi/cgi-bin/ssialias?htmlfid=WRL12345USEN}} and this rate of data creation is forecast to increase by 2025 up to $18 \times 10^{22}$ bytes per annum, the transfer of which with today's broadband technology would take 450 million years\footnote{The Economist, May 6th 2017 issue: The world's most valuable resource - Data and new rules of competition}. The main bulk of this `big data' is formed by records or ``digital footprints'' from various modes of modern-day communication such as mobile phone calls, social media postings, and online commercial activities. The availability of such data has opened up an unprecedentedly wide range of opportunities for computational analytics and modeling research on social systems in terms of their complex structure and dynamics~\citep{buchanan2008social}. In this data-driven approach one focusses on studying the complexities of social systems or networks at different structural and dynamical scales to understand ``how does microscopic translate to macroscopic''. In order to get insight into the {\em structure}, {\em function} and {\em behaviour} of social systems, we employ the methodologies of {\em data analysis} for making observations or discoveries, {\em modelling} for describing dynamical processes or plausible mechanisms governing the system and {\em simulation} for predicting the behaviour of the system. In this empirical research framework, which we call {\em Social Physics}, the broad perspective is about understanding the dynamics of human social behaviour in different contexts through phenomenological models. In constructing such models of complex social systems we are somewhat handicapped because unlike in physics we are not yet familiar with the ``underlying laws''. Then the purpose of the model becomes important. From physics we already know that any model does not come even close to capturing all the details of the system. Therefore, we have become accustomed to the idea that ``the model should be as simple as possible but not simpler'', yet we want the model to describe some of the salient features or behaviour of the real system, reasonably well. Thus in our model building we aim for tractability and clarity, by considering that `models are like maps' so that they are useful when they contain the details of interest and ignore others\footnote{A passage from Lewis Carroll's {\it Sylvie and Bruno Concluded} illustrates this point: “What do you consider the largest map that would be really useful?” “About six inches to the mile.” “Only six inches!” exclaimed Mein Herr. “We very soon got six yards to the mile. Then we tried a hundred yards to the mile. And then came the grandest idea of all! We actually made a map of the country, on the scale of a mile to the mile!” “Have you used it much?” I enquired. “It has never been spread out, yet,” said Mein Herr: “The farmers objected: they said it would cover the whole country, and shut out the sunlight! So now we use the country itself, as its own map, and I assure you it does nearly as well.”}. Therefore, we believe that the utility of simple models in describing the complexities of, for example, poorly understood ICT-based social systems, is very high. Simple models may, indeed, give deep insights into the social system in the same way that the simple Ising model provides useful understanding and quantitatively correct predictions on critical phenomena of real magnetic systems.

Over the past few years Social Physics has evolved into a multidisciplinary area of research interests focusing on human sociality embedded in social network and societal structures, with the help of data from various sources and its analytics. It is capable of revealing the structure and behavioural patterns of social systems at different scales from individual to societal level as well as capturing the long term evolution of the society~\citep{pentland2015social}. 
The concept of Social Physics is not new; it was first introduced by philosopher August Comte during the era of Industrial Revolution in the early 19th century with the view 
that the behaviour and functions of human societies could be explained in terms of underlying laws like in physics~\citep{ball2004critical}. In the first half of the 20th century there were pioneering contributions in this yet emerging area by scientists like G.~K. Zipf and J.~Q. Stewart~\citep{barnes2014big}. Since then the area has mainly evolved through the efforts of statistical physicists, especially through modelling ~\citep{stauffer2004introduction,galam2012sociophysics,sen2013sociophysics}. Now a full fledged data-driven paradigm, the Social Physics research borrows concepts and methodologies from social experiments, network science, game theory, theory of phase transitions and critical phenomena, automated data-collection systems, etc.  Overall, the Social Physics aims to answer scientific questions that would be instrumental in tackling challenges in the areas like, socio-political conflicts, organized crimes, human health, human migration, and the development and productivity at multiple scales from organizations to cities~\citep{san2012challenges,conte2012manifesto}.

Here we will review the work carried out in the spirit of Social Physics relying on data from mobile phone call detail records (CDRs), that includes, time stamps of voice calls and text messages (SMS messages). For the purpose of academic research such datasets could be released by mobile phone service providers with anonymity of the subscribers ensured~\citep{blondel2013mobile}. In another type of research with volunteers, the subjects have been allotted preprogrammed mobile phones to allow the collection of the data from the subjects into centralized servers for the time period of the study for the purpose of subsequent data analytics~\citep{eagle2006reality,stopczynski2014measuring}. One is referred to 
section~\ref{sociality} for further clarification on the nature of the data. Before such datasets and methods of data collection became available, research of human social behaviour relied mostly on questionnaire-based surveys and field observations typically on small number of individuals~\citep{zachary1977information}. In these survey-based studies the scope of social interactions is usually wide but based on individual recollection, raising the issues of subjectivity and how to measure the strength of social interaction, and with what scale. In contrast, the kind of data from mobile CDRs provide relatively narrow scope to social interaction while being based on measurements, for example in terms of the duration of a phone call between a pair of people. The real benefit of studies based on the mobile phone CDRs is that they have allowed making {\it in vivo} or {\it in situ} observations on large societal level populations over a wide range of spatial and temporal scales. Nevertheless, the mobile phone CDR based studies should be considered complementary to the questionnaire based survey studies. In this review We provide an overview of the work that includes the characterization and modeling of social dynamics at the level of individuals as well as the aggregate levels observed at different scales.   

A number of studies has already been conducted with mobile phone datasets from different perspectives including physics, communication engineering, transportation engineering, epidemiology, and sociology, to name a few. In this wide sense the current review is not an extensive one, and it is aligned to the approach of Social Physics. We will here focus on a set of selected topics, namely, human sociality (section~\ref{sociality}), human burstiness in time series of human communication (section~\ref{burstiness}), human mobility patters (section~\ref{mobility}), and the human daily and seasonal rhythms (section~\ref{circadian}). Readers wanting to learn more about related works as well as various other related topics may explore existing reviews, for example by \citet{blondel2015survey}, \citet{saramaki2015seconds}, and \citet{naboulsi2016large}. 

\section{Human sociality}
\label{sociality}

The study of human sociality from telecommunication data has over time developed into a unique field of research that provides insight into the micro-, meso- and macroscopic structure and dynamics of social networks of large-scale societal level populations. The paradigm is held in contrast to the method of surveys \citep{wasserman1994social} for collecting data on human interactions \citep{eagle2006reality,eagle2009inferring}. While, at the macroscopic level, the studies have primarily investigated the structural aspects of the social networks derived from the calling information aggregated over time windows \citep{onnela2007structure,onnela2007analysis,lambiotte2008geographical}, at the microscopic level the focus has been on ``egocentric networks'' to study the dynamics of closest relationships between egos (the individuals in focus) and their alters (the contacts of an individual) \citep{palchykov2012sex,miritello2013time,saramaki2014persistence,bhattacharya2016sex}. At the intermediate or mesoscopic level, social groups or communities and network motifs have been the objects of study \citep{palla2007quantifying,tibely2011communities,kovanen2011temporal,kovanen2013temporal,sekara2016fundamental}.

\begin{figure}[t]
\centering
\includegraphics[width=0.9\textwidth]{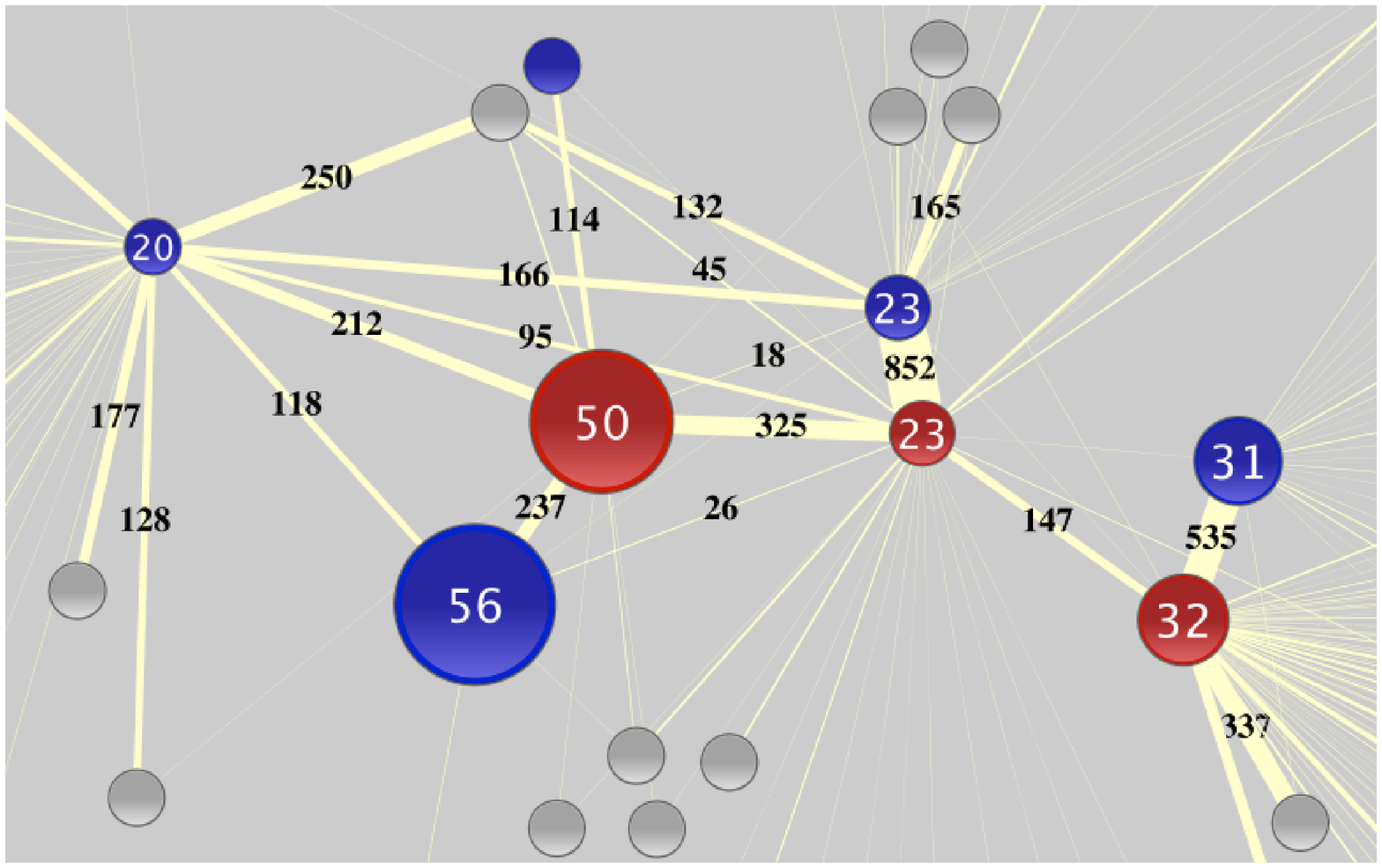}
\caption{A part of a network constructed from mobile phone call data. Blue circles correspond to male and red circles to female subscribers. The numbers inside a circle (and also its size) indicates the age of the subscriber. Grey circles correspond to subscribers whose gender and age information is not available in the data set. The numbers on a link (as well as its width) indicates the total number of calls between the connected pair of individuals over a period of seven months. From \citet{palchykov2012sex}.}
\label{egonet}
\end{figure}

A mobile phone CDR dataset comprises of call logs of egos or individuals with each log containing information of the timings of incoming and outgoing calls and of text messages i.e. SMSs. For calls, usually the information on the duration of the call is also included. The CDRs used for the purpose of research are anonymized, and the egos are distinguished by attributed surrogate keys. For a given ego and for a given communication event (a call or SMS) the corresponding alter key also appears in the log. The data acquired from mobile phone service providers may contain ancillary information on self-reported age, gender and residential postcode of the users. Using CDRs the social network is constructed by aggregating information within a time window (see Fig.~\ref{egonet}). For the studies in which volunteers are recruited, the investigators usually avail multiple modes of data collection. Apart from call logs, additional data could be collected from pre-installed software as well as from various sensor signals~\citep{eagle2006reality,karikoski2010measuring,stopczynski2014measuring,wang2014studentlife}. For example, the Bluetooth data from a volunteer collected at a certain instant could also indicate the presence of other Bluetooth devices, and hence other volunteers in the proximity~\citep{boonstra2017validation}. The studies of the latter kind usually also collect information from the participants in the form of questionnaires~\citep{saramaki2014persistence}. The data obtained from multiple sources usually helps in validating the results and dealing with the problems of missing data.

\subsection{Marcoscale and mesoscale patterns}

The first of its kind `Reality Mining' experiment was at the MIT by \citet{eagle2006reality} who conducted a study with 100 participants who were either studying or working at the institute and were allotted Nokia 6600 mobile phones. Using data on Bluetooth based proximity, the investigators could predict the type of relationship, namely, friend or office acquaintance with a high enough accuracy. \citet{onnela2007structure} made a large scale on networks using CDRs from a mobile phone service provider that had about $20\%$ of the market share in a European country. In the network two individuals were connected by a link if there had been a minimum of one reciprocated pair of phone calls between them over a period of 18 weeks. They termed this kind of a construct as a `mutual network' or as a proxy of a social network, and suggested that this method should eliminate a large number of one-way calls where the caller does not personally know the callee~\citep{onnela2007analysis}. Also, the weights of the links were attributed using the total calling duration between the ego-alter pairs. The final network that was analysed had about 4.6 million nodes and 7 million links. 

A major finding of \citet{onnela2007structure} was the confirmation of the {\it strength of weak ties} hypothesis~\citep{granovetter1973Strength}, according to which the weights of the links and the network topology were found to be correlated such that the intra-community links are stronger compared to the inter-community links. Also, by examining the percolation transition~\citep{stauffer1994Introduction} of the giant component containing $84\%$ of the nodes, they showed the importance of weak ties in maintaining the overall connectivity of the network. \citet{lambiotte2008geographical} conducted a large scale study on users based in Belgium with the mutualized network having 2.5 million nodes and 5.4 million links. Besides characterizing the topology of the network, they also showed the influence of language on the structure of the network (Belgium has native French and Flemish as well as bilingual speakers). 

The Belgian dataset was also used to investigate the community structure with a modularity optimization method, and the results showed that most communities are monolingual~\citep{blondel2008fast}. Similarly,~\citet{tibely2011communities} used the dataset studied by \citet{onnela2007structure} to obtain the community structure using three different community detection algorithms~\citep{lancichinetti2009community}. They compared the results from the different methods as well as illustrated the Granovetterian `weak ties hypothesis' from correlations between the link weights and community structure. \citet{palla2007quantifying} included the same dataset in their study of the evolution of community structure. They studied networks derived from the mobile phone calls and from co-authorship information in the Cornell University cond-mat eprint archive. Palla et al. showed the differences between the dynamics of small and large groups. 

\begin{figure}[t]
\centering
\includegraphics[width=0.8\linewidth]{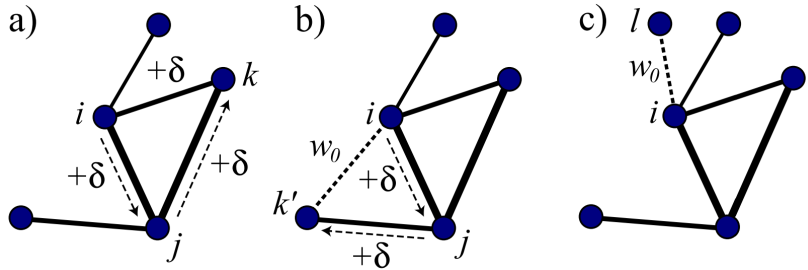}
\caption{Schematic diagram of the model by \cite{kumpula2007emergence}. 
(a): a weighted local search starts from $i$ and proceeds to $j$ and then to $k$, which is a also a neighbour of $i$. 
(b): the local search from $i$ ends to ${k'}$, 
which is not a neighbour of $i$. In this case link $w_{ik'}$ is set with probability $\ptri$.
(c): node $i$ creates a link randomly to a random node $l$ with probability $\pr$.
In the cases of (a) and (b) the involved links weights are increased by $\delta$. 
}
\label{fig:algoritmi}
\end{figure}

\begin{figure}
\centering
\includegraphics[width=0.9\linewidth]{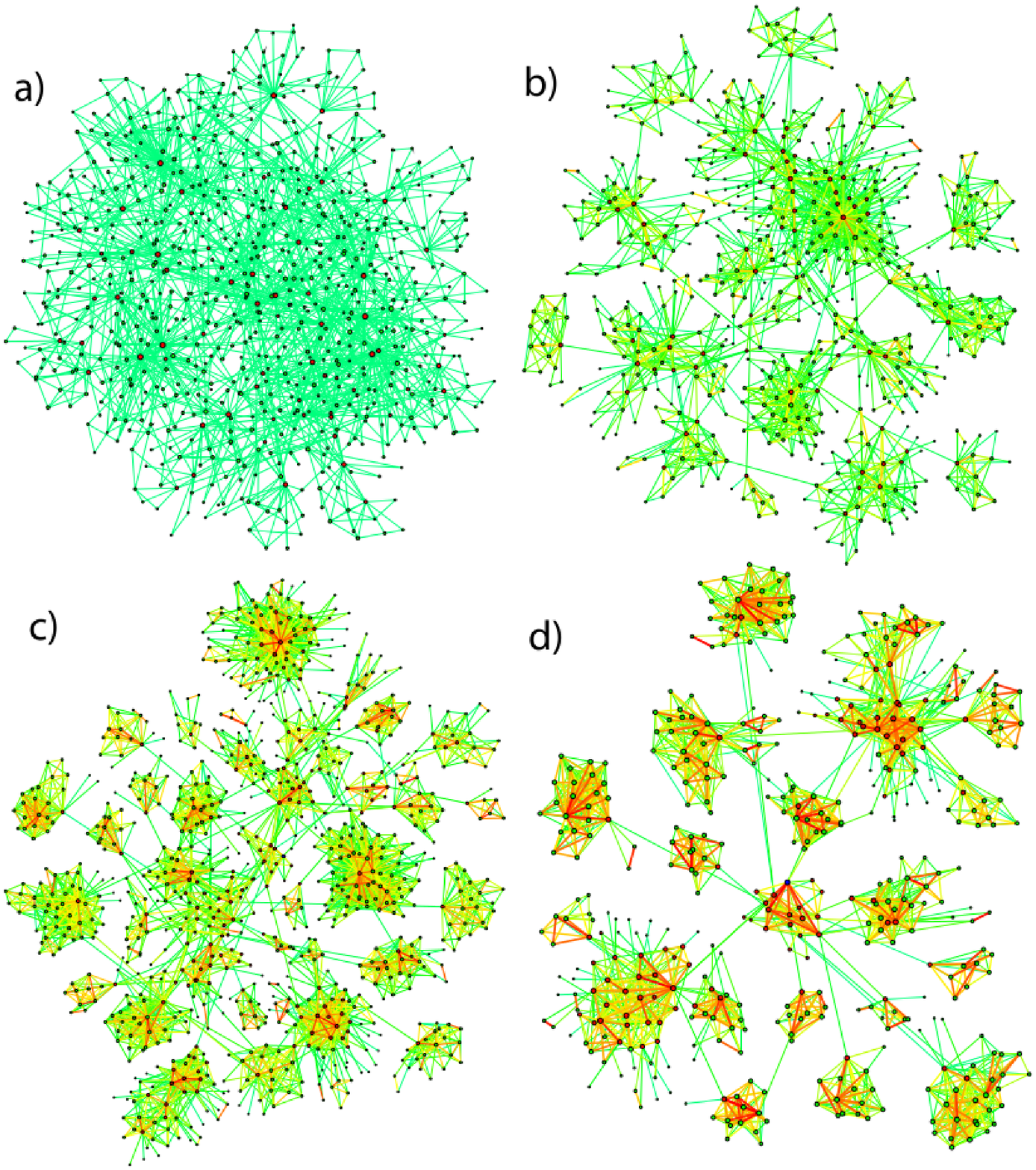}
\caption{Simulated networks in the model by \cite{kumpula2007emergence}, with (a) $\delta=0$, (b) $\delta=0.1$, (c) $\delta=0.5$, and (d) $\delta=1$. Link colours change from green (weak links) to yellow and red (strong links). With increase in the value of $\delta$, community structure starts appearing.}
\label{model2.eps}
\end{figure}

The Granovetterian hypothesis was also illustrated in the context temporal networks by \citet{kovanen2013temporal}, where the authors carried out a census on temporal motifs in the mobile phone data (see also \citet{kovanen2011temporal}).  A more recent study of meso-scale patterns using the Copenhagen Network Study (CNS) dataset proposed a general framework for an evolving social network~\citep{sekara2016fundamental}. The dataset involving around 1000 freshman University students, contained information of calls, SMSs, proximity via Bluetooth, geo-location, and social media contacts in addition to general demographic information. Sekara et al. proposed to identify stable cores in dynamic networks, representing meetings between individuals having different social contexts like work and recreation. They also suggested that the concept of cores is better suited to the description of social groups in time varying networks as compared to the other community detection algorithms.

As one of the first examples of an approach towards modelling meso- and macroscopic structures of a human mobile communication type social network, we describe a model presented by~\citet{kumpula2007emergence}. This model describes the processes for individuals getting acquainted with each other leading in turn to the formation of locally and globally complex weighted social network structures. In the model one considers a network with a fixed  number of $N$ nodes (or individuals) where the time evolution of the structure proceeds with the following three rules. (i) The cyclic closure or local attachment rule: In a time interval $\dt$ each node or individual having at least one neighbour starts a weighted local search for new friends, see Fig.~\ref{fig:algoritmi}(a,b). The node $i$ chooses one of its neighbouring node $j$ with probability $w_{ij}/s_i$, where $w_{ij}$ represents the weight of the link connecting $i$ with $j$ and $s_i=\sum_j w_{ij}$ is the strength of node $i$. If the chosen node $j$ has other neighbours $k$ (apart from $i$), it chooses one of them with probability $w_{jk}/(s_j - w_{ij})$ implying that the search favours strong links. If there is no connection between $i$ and $k$, it will be established with probability $\ptri \dt$ such that $w_{ik}=w_0$. If the link exists, then its weight will be increased by a certain amount $\delta$. In addition, both $w_{ij}$ and $w_{jk}$ are increased by $\delta$. This kind of cyclic linking simulates the mechanism where ``friend of a friend will also be friend''. The focal closure or global attachment rule is as follows: If a node has no links, it will create a link of weight $\wo$ with probability $\pr \dt$  to a random node, as depicted in Fig.~\ref{fig:algoritmi}(c). This mechanism is to establish a new link outside the immediate neighbourhood of the chosen node.  (iii) The node deletion rule: With probability $\pd \dt$  an existing node with all its links are removed from the system, and new node is introduced to maintain fixed system size. The parameter $\delta$ is responsible for the time-dependent development of the weights of the network. When $\delta=0$, one obtains unweighted networks without apparent community structure. With the increase in $\delta$, emerging triangles become the nuclei for community formation. Networks simulated for different values of $\delta$  are shown in Fig.~\ref{model2.eps} ($\dt=1$, $\wo=1$, $\pd=10^{-3}$, and $\pr=5\times10^{-4}$). For larger $\delta$ values one sees the formation of a community structure very similar to that observed in the studies of the mobile phone dataset \citep{onnela2007structure,onnela2007analysis}.

\subsection{Egocentric networks}

While the investigations at the macroscopic and mesoscopic scales shed light on the aspects of social-like societal structure and the dynamics of group formation, the egocentric networks focus on the structure and dynamics of social ties within the immediate social neighbourhood of an individual. Different research questions can be addressed depending on the nature of data at hand. For a dataset with some ($\sim 30$) student volunteers being observed over a period of several months, the study of it shows how for a given ego the alters may change their ranks over time~\citep{saramaki2014persistence}. On the other hand if the dataset comprises of millions of subscribers whose age and gender information is available, their social behaviour can be analysed as a function of these variables, even if the dataset only spans over a month~\citep{dong2014inferring}.  

In this context \citet{palchykov2012sex} were first to rank the set of alters by the total number of calls and they examined the gender and age of the top ranked alters as a function of the ego's age. Based on previous studies they assumed that the communication intensity should be a marker of emotional closeness~\citep{roberts2009exploring}. They observed that, while the primary focus of males of all ages is likely to be on the opposite gender (or spouse), for older females there is a shift from the spouse to younger females, who can be assumed to be the daughters on the basis of the age difference. Palchykov et al. mentioned that the observation was in line with the {\it grandmothering} hypothesis~\citep{hawkes2004human}. \citet{palchykov2013close} also proposed a statistical method for ranking the set of alters based on the different relevant quantities (i.e. number of calls, duration of calling, number of SMSs, etc.).  

A question that has been addressed in a number of studies is how an ego distributes its total time of communication among its alters. Using a large dataset,~\citet{miritello2013time} calculated a measure for disparity and found it to be negatively dependent on the number of alters, suggesting different strategies of communication between the users who have large numbers of contacts and those who have few. In general, all the users were found to distribute their limited time very unevenly across their contacts. Interestingly, the average communication time per alter, though being positively dependent on the number of alters, started to decrease when the number of alters increased beyond 40. Miritello et al. suggested that this effect was a partial reflection of the social brain hypothesis ~\citep{dunbar1993coevolution}, according to which there is a cognitive limit in the number of social contacts an individual can maintain, i.e. (the {\it Dunbar number} $\sim 150$). In another study,~\citet{miritello2013limited} characterized the social strategy for communication in a space spanned by the communication capacity and communication activity of individuals. They identified two kinds of extreme strategies, namely, {\em social keeping} in the case of which the individuals keep a very stable focus on selected contacts, and, {\em social exploring} when the individuals activate and deactivate social ties at a high rate. \citet{saramaki2014persistence} showed that even in the presence of a turnover occurring in the egocentric network, the pattern in which an ego allocates its communication across the alters is stable. Also, that this pattern is usually the characteristic of an individual and could be considered as a ``social signature'' of that person. In a recent study, \citet{centellegher2017personality} have explored the connections between the social signatures obtained from mobile phone calling and the individual personality traits of the participants obtained using Big five model surveys\footnote{\url{https://en.wikipedia.org/wiki/Big\_Five\_personality\_traits}}. They observed the extroverts to have slightly lower temporal persistence of their social signatures, as compared to the introverts.

\citet{bhattacharya2016sex} investigated further the aspect social focus using a large dataset~\citep{onnela2007structure,palchykov2012sex}. They found that younger individuals have more contacts and, among them, males more than females. In addition, they observed a steady decrease in the number of contacts with different rates for males and females, such that there is a reversal in the number of contacts around the age of late 30s. They suggested that this pattern can be attributed to the difference in reproductive investments made by the two genders. Bhattacharya et al. also analysed the gender-based inequality or disparity in social investment patterns by measuring the Gini coefficient and found the women and older people to have slightly larger values. The study was further extended by \citet{david2016communication,david2017peer}, where the authors characterized different types of social ties (friendship, spousal, kinship) based on age and gender information of the alter in conjunction with the ego. They showed how people have the tendency to rearrange their social lives as they grow older. 

\begin{figure}
\centering
\includegraphics[width=0.9\textwidth]{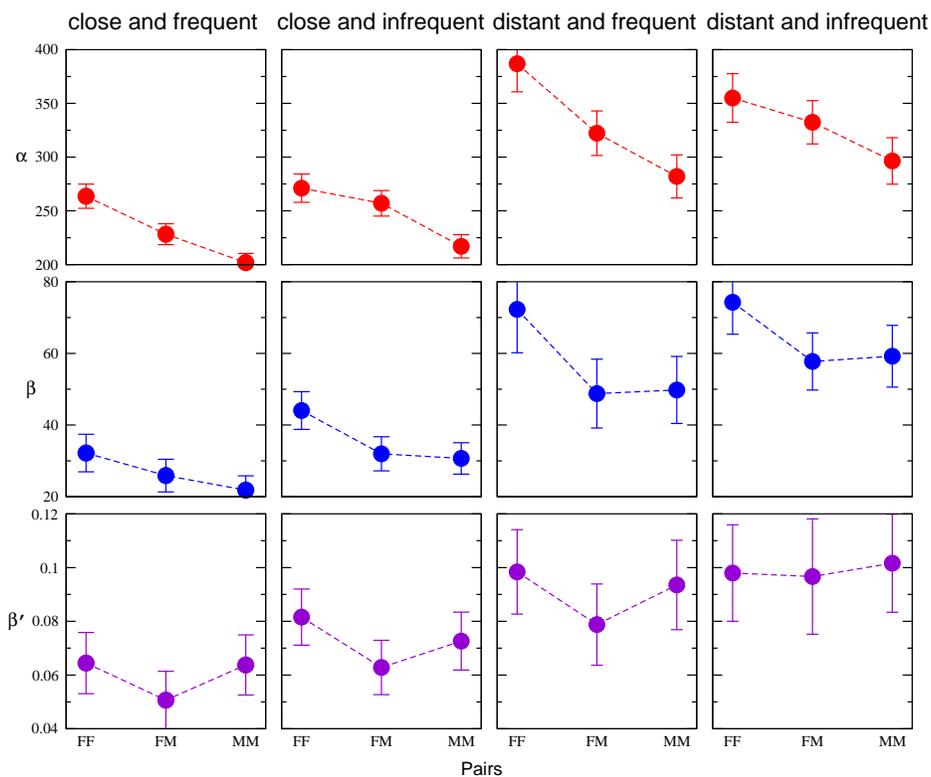}
\caption{Quantifying the dependence of the duration of the succeeding call ($T$, measured in seconds) on the inter-call gap ($\tau$, measured in number of days) for pairs of regularly communicating individuals.  The coefficients resulting from the regression: $T=\beta\log\tau+\alpha$, are shown. $\beta'$ is the coefficient when both the duration and the gap are scaled by their respective averages. A broad distinction into four groups (as indicated on the top of the columns) is done based on whether for a given pair the distance between their locations is less or greater than 30 km (i.e., geographically ``close'' or ``distant'', respectively) which being the spatial extension of large cities; and whether average gap, $\langle\tau\rangle$ is less or greater than 12 days (i.e., communication is ``frequent'' or ``infrequent'') which being the most probable inter-call gap. A finer classification is made based on the gender of the individuals as indicated along the horizontal axis (FF: female-female, MM: male-male, FM: mixed). Pairs are chosen irrespective of their age. The dashed line is a guide to the eye. From \citet{bhattacharya2017absence}.}
\label{maintenance}
\end{figure}

The maintenance of social relationships, in particular, friendships is another important context of egocentric networks. \citet{bhattacharya2017absence} have demonstrated the maintenance of social relationships by examining the dependence of call durations on inter-call gaps (inter-event times) for pairs of individuals who regularly call each other. They found a logarithmic increase in call duration ($T$, measured in seconds) with an increase in the inter-call gap ($\tau$, measured in number of days): $T=\beta\log\tau+\alpha$. The authors suggested that the increase in the duration of the succeeding call after a long gap (time gap with the previous call) could serve as an act of relationship repair. In a sense this finding can also be considered as reflecting the maintenance of the strength of Granovetterian links. The authors have also examined the effect with the set of alters categorized into different subsets (Fig.~\ref{maintenance}). \citet{navarro2017temporal} have investigated the persistence of ties in terms of a prediction model based on the intensity, intimacy, and structural and temporal features of social interaction between a pair of individuals. These features included variables, like the total number of calls, age difference, topological overlap~\citep{onnela2007structure}, reciprocity~\citep{chawla2011weighted}, and coefficient of variation in inter-call gap. Their analysis showed that temporal features are effective and efficient predictors, and that bursty communication is correlated with decay (see section~\ref{burstiness}). In slightly different context, \citet{tabourier2016predicting} investigated  the problem of link prediction~\citep{lu2011link} for egocentric networks. They constructed a machine learning scheme using a mixture of supervised and unsupervised methods that, for a given ego, would predict the presence or absence of a link between a pair of its alters.


\section{Burstiness in the time series of human communication}
\label{burstiness}
Having looked at patterns observed at the level of information aggregated from calls, we now focus on the topic of `burstiness' that becomes relevant when the times series of the calling activity of the service subscribers are investigated. The dynamics of a system is typically termed as bursty when there are short periods of heightened activity followed by longer periods of inactivity \citep{goh2008burstiness,barabasi2010bursts,karsai2018bursty}. This type of dynamics has been observed in various kinds of natural and man-made systems where the dynamics can be characterized in terms of a time series.  Some examples of them are earthquakes~\citep{bak2002unified}, neuronal firing~\citep{grace1984control}, biological evolution~\citep{uyeda2011million}, email patterns~\citep{eckmann2004Entropy}, and mobile phone calls~\citep{karsai2012Correlated}. Systems displaying bursty dynamics are typically differentiated from systems displaying Poisson processes \citep{grimmett2001probability}. In the latter case the rate at which events occur is a constant that is independent of time, and the resulting probability distribution for the time between a pair of successive events (inter-event time) is an exponential. Burstiness in time series is found to result due to the departure of the distribution of inter-event time from being an exponential, as well due to the presence of correlations between inter-event times. The time series found in nature and society where burstiness is observed are often described by the probability distribution ($P(\tau)$) of inter-event times ($\tau$) that is a power law with an exponential cut-off: $P(\tau)\sim\tau^{-\alpha}\exp(\tau/\tau_c)$, where $\alpha>0$, and $\tau_c$ is a time scale originating from the finiteness of the time series \citep{karsai2012universal}. Other commonly used measures to characterize the heterogeneity and correlations in the inter-event times include the burstiness parameter and the memory coefficient~\citep{goh2008burstiness}. 

\begin{figure}
\centering
\includegraphics[width=0.9\textwidth]{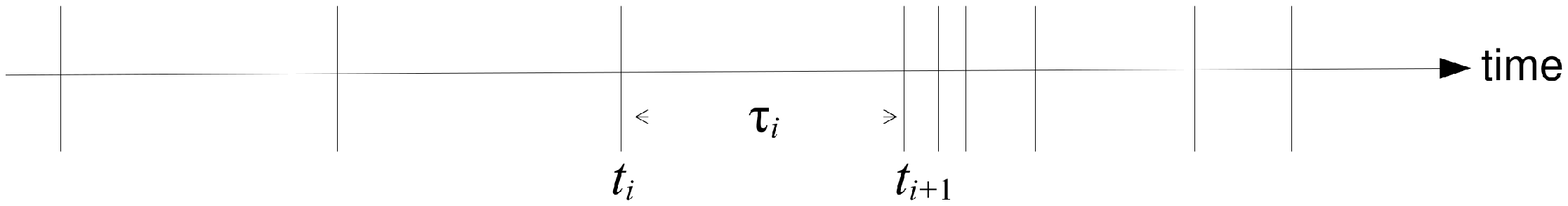}
\caption{Schematic diagram of a time series of events. Each vertical line indicates the timing of an outgoing call (event) from a given individual. The inter-event times ($\tau$) are calculated from the differences in the timings of consecutive events, with $\tau_i=t_{i+1}-t_i$. The distribution of inter-event times ($P(\tau)$) for a bursty time series is usually heavy-tailed.}
\label{time-series}
\end{figure}

For mobile phone-based interactions between people the time series' are constructed from the the CDRs of individual users by marking the timings of the events, namely, calls and SMSs. The calls are usually of finite duration which is rather discarded, and the starting time is considered (see Fig.~\ref{time-series}). Also, for calls as well as SMSs the outgoing and incoming events may be separately analyzed \cite{aoki2016input}. A number of empirical results support the power law form for the inter-event distribution $P(\tau)$, where the exponent $\alpha$ is generally found to depend on whether it is from calls or from SMSs. \citet{karsai2012universal} used the data from a European operator \citep{onnela2007structure} to show that $\alpha\approx 0.7$ for calls and $\alpha\approx 1.0$ for SMSs. \citet{aoki2016input} also used the data from a European operator \citep{tabourier2016predicting} and found $\alpha$ to be $1.2$ and $1.4$ for calls and SMSs, respectively. Using a Chinese dataset, \citet{jiang2013calling} found that although the aggregate $P(\tau)$ follows a power law, a majority of individual users show Weibull distributions for inter-event times. For individuals with power-law the values of $\alpha$ varied from $1.5$ to $2.6$. The authors observed that anomalous and extreme calling patterns were associated with power laws, which could link the users to robot-based calls, telecom frauds, or telephone sales.

In general, the heavy-tailed $P(\tau)$ has been observed in several different studies with calls and SMSs~\citep{candia2008uncovering,miritello2011dynamical,schneider2013unravelling}. Interestingly, bimodal distributions of inter-event times have also been observed in SMS datasets~\citep{wu2010evidence} such that the distributions are power-law for $\tau<\tau_0$, and exponential for $\tau>\tau_0$. \citet{wu2010evidence} reproduced the observations using a model of interacting queues for individuals, and showed that the time scale $\tau_0$ could be related to a parameter called `processing time' in the model.  

Motivated by the observations from mobile call and SMS data, and communication data in general, there have been several attempts to reproduce the features related to burstiness using agent based models. \citet{karsai2018bursty} divide the models into three broad categories: (a) models of individual activity~\citep{Barabasi2005Origin,Vazquez2006Modeling,Vazquez2007Impact2,Jo2015Correlated}, (b) models of link activity~\citep{Oliveira2009Impact,wu2010evidence}, and (c) network models of bursty agents \citep{Goetz2009Modeling,stehle2010dynamical,jo2011emergence}. Apart from the above type of models, there have been investigations of dynamical processes using dynamical substrates or temporal networks such as a mobile phone call network that has burstiness inherent to its structure~\citep{holme2012temporal}. Processes that have been studied include epidemic spreading~\citep{karsai2011small,kivela2012multiscale}, random walks~\citep{speidel2015steady} and threshold-driven contagion models~\citep{takaguchi2013bursty,backlund2014effects}.

Here we briefly describe the model by \citet{jo2011emergence} that is inspired by the model og \citet{kumpula2007emergence} with the cyclic and focal closure mechanisms for network formation (see section~\ref{sociality}) and incorporates the concept of queues. In this model, $N$ agents are each allotted a list containing tasks of two types, $I$ and $O$. The priorities of each task is randomly chosen from a uniform distribution. At each time step $t$, every node selects the task with the highest priority. If it is an $I$-task, the actual node $i$ selects a target node for interaction either (i) from the whole population with probability $p_{_{GA}}$, or (ii) from its next nearest neighbours with probability $p_{_{LA}}$, or (iii) from its neighbours with probability $1-p_{_{GA}}-p_{_{LA}}$ weighted by their link weight $w_{ij}$. The next nearest neighbour $j$ of the node $i$ is defined as a node that satisfies $\{ t_{ik},t_{jk} \} =\{ t-2, t-1 \}$ with an intermediate node $k$, implying that $i$ and $k$ interacted at time $t-2$, and $k$ and $j$ interacted at time $t-1$. In the first two cases, (i) and (ii), links with unit weights are created between the interacting nodes. In case of (iii) the process is reinforcement in terms of an increment of  existing weights. After an interaction the priority of $I$-task of node $i$ is updated. In addition, at each time step each node can forget all of its existing connections with a probability $p_{_{ML}}$, i.e., having a memory loss, to become isolated. By measuring the inter-event times between two consecutive $I$-tasks of a given node the system exhibits a broad inter-event time distribution with an exponential cutoff. In addition, the emerging network structure shows several realistic features similar to the model by \citet{kumpula2007emergence}.

\section{Human mobility patterns}
\label{mobility}
\begin{figure}[h!]
\centering
\includegraphics[width=0.75\textwidth]{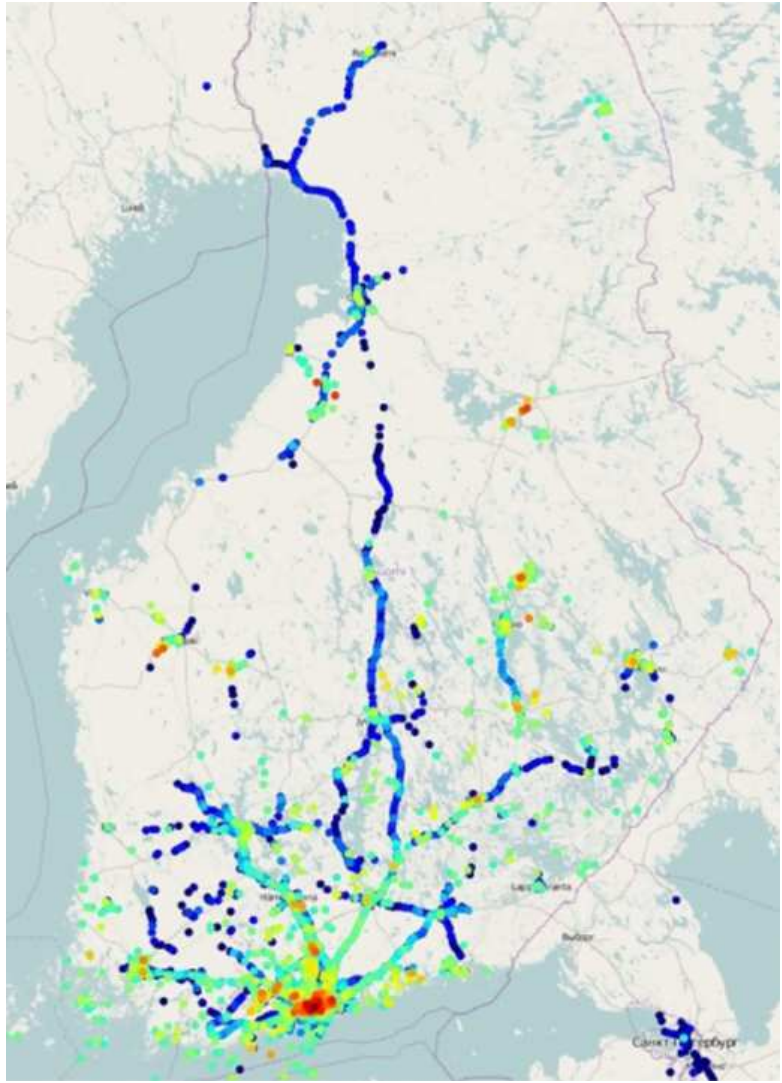}
\caption{Positional stamps of mobile phone communication of the subjects in the Otasizzle project~\citep{karikoski2010measuring} at cell towers located around Finland for a period spanning 16 months. There were around 180 subjects, who were either students or staff members of the Aalto University, Finland, and were using Nokia Symbian smartphones with pre-installed software. The stamps got recorded every half an hour, or whenever a phone got connected to a new cell tower. Each cell tower could be located with a unique pair of latitude and longitude degrees. The higher the frequency of usage, the warmer the colour. The spatial trajectory of an individual user could be determined by following the sequence of the towers used. From~\citet{jo2012spatiotemporal}.}
\label{mobility-fig}
\end{figure}

The idea of random motion of material particles while interacting with complex environments has long interested physicists \citep{bouchaud1990anomalous,havlin2002diffusion}. With access to imaging technologies and other data logging methods a whole research area has emerged for studying the movement patters of organisms from microbial scales to the scales of primates and other mammalian species \citep{ramos2004levy,viswanathan2011physics}. A question underlying such studies has been how to model the paths of individual entities that can decide for themselves. A popular candidate has been the L\'evy flight process in which the move lengths follow an inverse power law \citep{viswanathan1999optimizing}. Addressing similar questions in the case of humans have been possible with the advent of mobile phones that have facilitated to infer the trajectories of users. For billing purposes the service providers usually record the cell tower that is involved in routing a call. Therefore, when a user initiates or receives a call the location of the associated tower is known in terms of latitude and longitude (see Fig.~\ref{mobility-fig} where the data from the Otasizzle project is visualized). In certain studies, data from GPS and WiFi routers have allowed the recording of trajectories with finer resolution, for example, the researchers in the CNS~\citep{cuttone2018understanding} achieve a median spatial resolution of 20 meters by sampling every 15 minutes. Large-scale studies based of such spatio-temporal data have provided insights into the mobility patterns of humans. The investigations have addressed diverse topics ranging from the statistical properties of trajectories to predictive models and to characterization of urban landscapes.

\citet{gonzalez2008understanding} tested the L\'evy flights hypothesis on humans using two different data sets, the larger of which comprised of 100,000 individuals. They observed that with each user a characteristic `radius of gyration' could be associated, which served as a bound on individual trajectories. Using scaling and data collapse they showed that travel patterns of a users could be described by L\'evy flights for walk lengths below a radius of gyration, and extremely large displacements that would be predicted by a L\'evy flights hypothesis were rather absent. \citet{gonzalez2008understanding} also made the general observation that human trajectories are highly regular and are characterized by few highly frequent locations like home and workplace. Also, as a result of a heavy tailed distribution of radius of gyration within the population, the aggregated distribution of the move length of all users would still be a power law. These ideas were followed up in a work by \citet{song2010modelling} who provided a microscopic model laying stress on the  aspects of exploration of space and propensity to return to already visited locations. Describing the spatio-temporal trajectories of 90 thousand users in a countrywide dataset in terms of a weighted and directed network, \citet{bagrow2012mesoscopic} concluded that the daily trajectory of most of the users were dominated by 5--20 frequently visited locations or ``habitats''. In general, spatio-temporal motifs in the daily patterns of commuting individuals have also been investigated and modelled~\citep{schneider2013unravelling,jiang2016timegeo}. \cite{alessandretti2016evidence} have shown that a number of locations an individual visits regularly is conserved over time. The authors have suggested that this conserved number could be an outcome from the limited capacity of human cognition similar to the Dunbar number \citep{dunbar1993coevolution}. In an exploratory analysis of a dataset from Portugal, \citep{csaji2013exploring} showed how the locations of home and office can indeed be ascertained. They further used these locations to estimate the number of commuting trips ($x_{ij}$) between two counties $i$ and $j$, and modelled this quantity by using a gravity law. 

The gravity law having a form similar to Newton's law of attraction states that, $x_{ij}\propto m_i^\alpha n_j^\beta / d_{ij}^\delta$, where, $x_{ij}$ is a measure of social interaction between $i$ and $j$, with $m_i$ and $n_j$ being the 
populations of $i$ and $j$, respectively, and $d_{ij}$ is the distance between the geographical centres of the counties \citep{carrothers1956historical,carey1867principles,balcan2009multiscale,anderson2011gravity}. The parameters $\alpha$, $\beta$ and $\delta$ are usually estimated from the data. Studies considering $x_{ij}$ as the aggregated intensity of mobile phone communication have yielded different values of the parameters depending on the geographical origin of the dataset \citep{krings2009urban,lambiotte2008geographical,onnela2011geographic}. As a development over the gravity model, \cite{simini2012universal} proposed a `radiation model' for the mobility and migration between places in the framework of emission and absorption of particles from the sources and targets. Apart from modelling the commuting fluxes, they used the same dataset as \citet{gonzalez2008understanding} to model the volume of phone calls between municipalities. In a recent work, \citet{grauwin2017identifying} introduced a new model for the spatial organization of the population in a country based on the concept of a hierarchical distance. Analysing calling data from multiple countries, the authors observed a systematic decrease of communication flux induced by regional borders within countries. While the gravity and the radiation were unable to capture this effect, the new model produced better predictions due to the use of a hierarchical distance replacing the continuum geographical distance. Interestingly, \citet{louail2014mobile} studied the spatial organization of population in cities using aggregated CDRs providing the number of unique individuals using a given cell tower for each hour of the day. By studying the separation between the tower positions weighted by the density of users as function time during a typical day, the authors could discern between the patterns found for monocentric and polycentric cities.

Predictability and benchmarking of mobility patterns of individuals has been another important track. In this respect the concept of Shannon entropy has been a powerful tool. In the Reality Mining study, \citet{eagle2006reality} observed that people whose patterns yield low entropies should be easier to predict, and vice versa. In a seminal work, \citet{song2010limits} utilized three expressions based on the Shannon entropy to test the extent to which human mobility could be predicted. For example, an individual visiting $N$ locations could be characterized by a temporal-uncorrelated entropy, $S=-\sum_{i=1,N}p_i\log_2p_i$, where $p_i$ is the probability that a location $i$ is visited by the individual\footnote{Interestingly, in the expression for Shannon entropy $N$ could denote the number of alters of an ego such that $p_i$ is the fraction of calls between the ego and an alter $i$. This kind of expression has been used, for example by \citet{saramaki2014persistence}. In a study to investigate gender differences, \citet{psylla2017role} used entropy expressions in both the contexts of mobility as well as sociality.}. Additionally, they used a random, and a correlated entropy that took into account the order in which the locations were visited. Using these entropy expressions the authors employed Fano's inequality\footnote{\url{https://en.wikipedia.org/wiki/Fano\%27s\_inequality}} to estimate the upper bound of predictability that algorithms can achieve. The findings revealed that irrespective of the background of the individuals investigated, there was a high degree of regularity in the trajectories, with the average of the maximum predictability having a value around 93\%. \citet{lu2012predictability} used this framework with the data from a mobile phone users who were displaced following an earthquake in Haiti, and showed that predictability constructed from historical records of the individuals was quite high. The authors suggested that the knowledge about the possible future locations of people displaced by natural disasters would be useful in planning and dispensing of relief. 

Using data from three cities in two industrialized countries \citet{toole2015coupling} explored the links between social ties and  mobility. They constructed location vectors of individuals by counting visits to different locations in a city (number of entries in the vector), and subsequently obtained cosine similarities between pairs of individuals who were communicating. The results showed that the mobility pattern of an individual is, in general, similar to the latter's social contacts. Moreover, the stronger the social tie as observed from communication, the more was the similarity in mobility. In the study by \citet{deville2016scaling}, the authors used scaling theory and universality to obtain a scaling relationship between mobility and communication. They were able to connect the exponents of the two empirically observed power laws characterizing the distributions of aggregated move lengths and geodesic distances (as well as density dependent distances) between the communicating pairs. To establish the relation they showed that a mediating scaling relationship exists between fluxes of mobility (number of trips) and fluxes of communication (number of calls) between two locations.

More recently, using the CNS dataset,~\citet{mollgaard2017correlations} have shown that mobility as well as other types of behaviour of individuals can be predicted with high precision based on patterns obtained at the population level. They base their conclusion on a logistic regression model using information on calls, SMS, proximity via Bluetooth and mobility from GPS. Using the same dataset~\citet{cuttone2018understanding}, discussed the contrasting ideas of predicting a visit to an already visited location against predicting exploration of newer locations. In addition, they discussed the effect of spatial and temporal resolutions on predictive power. The related issue of sampling has been addressed in a recent study by~\citet{gallotti2018tracking}. The authors constructed a one dimensional model of alternating moves and rests, with sampling at regular intervals, and showed analytically that the fraction of correctly sampled moves can be 51\% at maximum. They applied the model to empirical datsets to show that the fraction decreases even further.   

\begin{figure}
\centering
\includegraphics[width=0.9\textwidth]{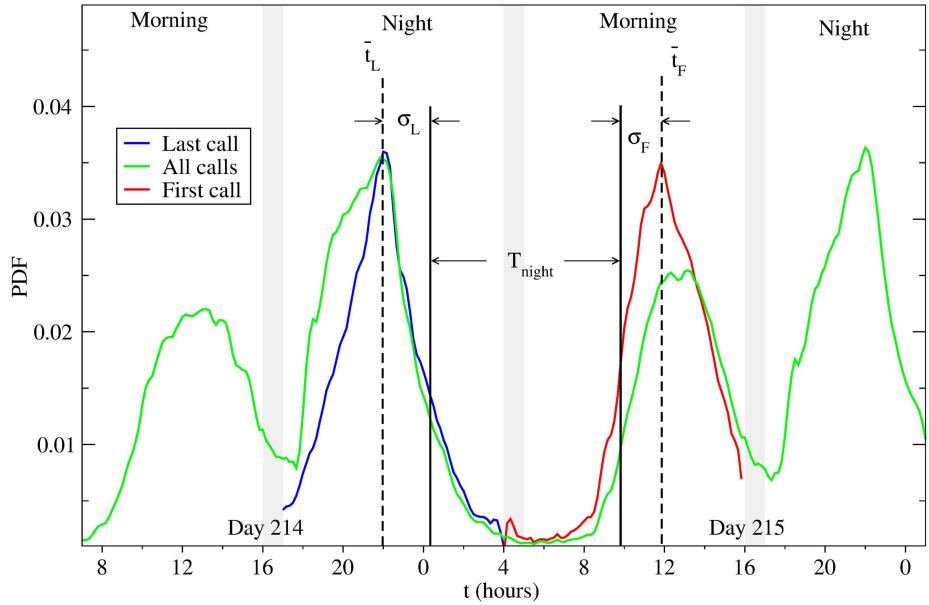}
\caption{Probability distribution functions (PDF) of finding an outgoing call at a time $t$ of the day in a city, for a pair of consecutive days in a given year. (Green) Distribution when all calls are included. (Red) Distribution when only the last call (L) at night is included (between 5:00 pm and 4:00 am next day).  (Blue) Distribution when only the first call (F) of the day is included (between 5:00 am and 4:00 pm). The respective mean times, $\overline{t}_L$ and $\overline{t}_F$, and the standard deviations $\sigma_L$ and $\sigma_F$ are calculated. A period of low calling activity is defined as the region bounded by $\overline{t}_L$ and $\overline{t}_F$, and its width $T_{night}$ is calculated as the time interval between $\overline{t}_L+\sigma_L$ and $\overline{t}_F-\sigma_F$. Interestingly, the onset and length of $T_{night}$ change along the seasons, in middle of February (day 46) $T_{night}$ is around $10.5$ hours, whilst in the early August (day 214) it is $9.5$ hours. From \citet{monsivais2017tracking}.}
\label{daily-pattern}
\end{figure}

\section{Daily and seasonal rhythms}
\label{circadian}
In section~\ref{burstiness} we focused on the temporal heterogeneity in people's social patters. Here we discuss the periodic nature of the temporal signals obtained from the CDRs of individuals \citep{jo2012circadian}. Remarkably, by precisely examining the daily, weekly, and seasonal patterns it is possible to gather insights into the different factors influencing human behaviour and health \citep{cornet2017systematic}. This is demonstrated very clearly by the Fig.~\ref{daily-pattern} where the authors~\citet{monsivais2017seasonal} show the daily pattern for outgoing calls in a city, from which they calculate a characteristic time duration that provides an upper bound for the  
sleeping time of the population. 

Traditionally, the field of chronobiology had been involved in the science of circadian rhythms - a roughly 24 hour cycle that is found in the biological processes of living organisms \citep{aschoff1965circadian,vitaterna2001overview}. Notably, the Nobel Prize in Physiology or Medicine of 2017 was awarded for the discovery of molecular mechanisms behind these rhythms\footnote{https://www.nobelprize.org/nobel\_prizes/medicine//medicine/laureates/2017/}. For humans, a circadian cycle encompass oscillations in different physiological processes, like blood pressure and body temperature, along with the alternating phases of sleep and wakefulness. It is known that the ``clocks'' that are internal to body, and are responsible for maintaining the rhythm are synchronized or entrained to various external factors of which the day-night cycle is likely the most well-understood one. Other factors like social interactions and environmental conditions are also known to be important \citep{grandin2006social,yetish2015natural}. The major goals for understanding the circadian rhythm and sleep or overall daily resting behaviour include characterizing populations with respect to physiological factors like gender and age, and at the same time constructing benchmarks for monitoring the health of individuals. The study on human subjects is usually conducted through questionnaires and wearable sensors \citep{ancoli2003role,roenneberg2015human}. In this regard, the daily records of usage of mobile and smart phones appears to provide a complimentary, unobtrusive and viable alternative  
\citep{choe2011opportunities,cornet2017systematic}. 

In the studies on sleep-wake cycles individuals are usually classified as morning-type and evening-type, based on the tendency to wake up early and the tendency to sleep late, respectively \citep{roenneberg2003life}. It is also known that due to busy work schedules evening-type individuals sleep less on weekdays \citep{roenneberg2007epidemiology}. As a compensatory effect the duration of sleep for these individuals gets stretched on weekends. Using the smartphone screen on-off sensor data from 9 volunteers collected over a period of 97 days,  \citet{abdullah2014towards} demonstrated this effect by quantifying ``sleep-debt''. They considered the screen on-off as an indicator of activity of the users. There have been other investigations of sleep, circadian rhythm, and mental health in general, based on smartphone data not restricted to screen on-off times but also including data from other sensors like accelerometer, microphone and GPS \citep{bai2012will,chen2013unobtrusive,ben2015next,abdullah2016automatic, saeb2016relationship}.

\begin{figure}
\centering
\includegraphics[width=0.9\textwidth]{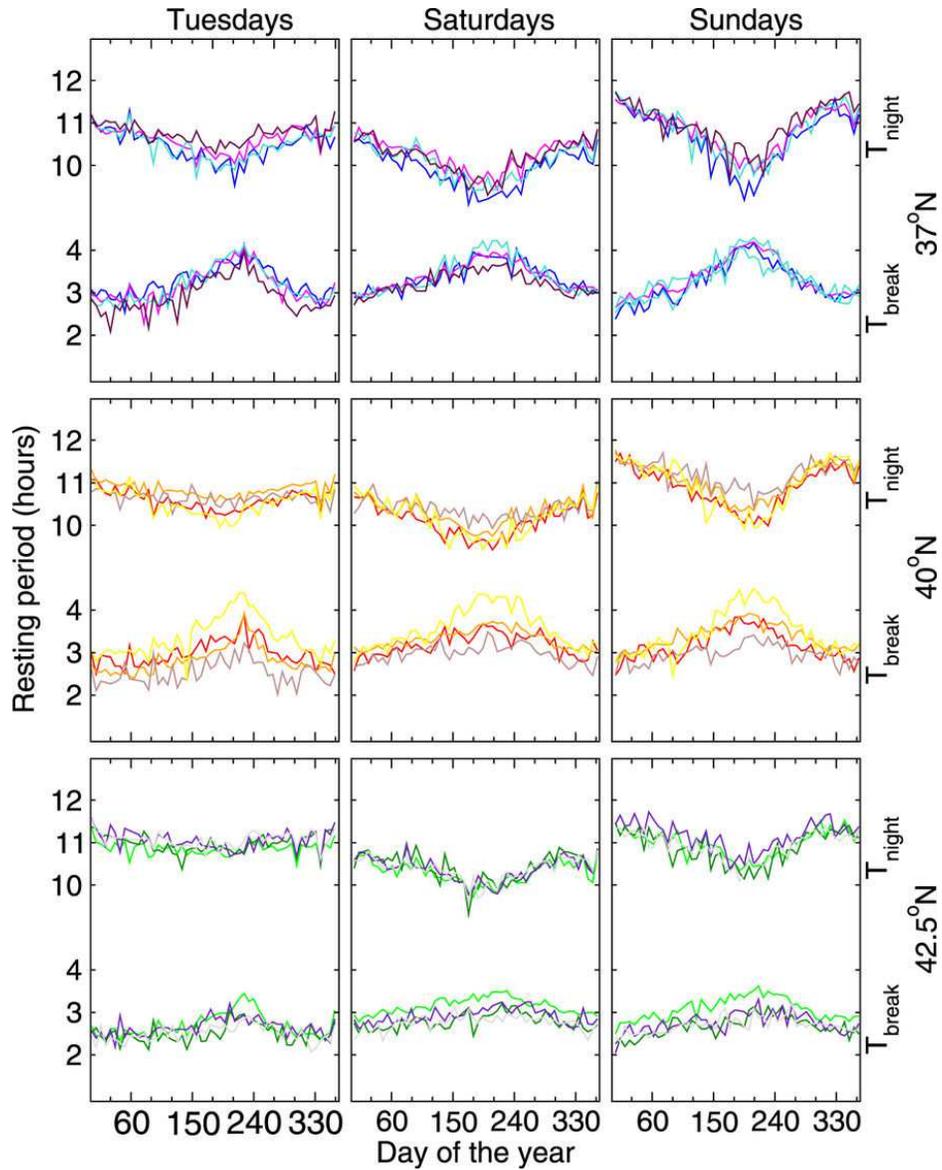}
\caption{Periods of low calling activity or resting periods as measured from intra-day distributions of calls (Fig.~\ref{daily-pattern}): $T_{break}$ (afternoon) and $T_{night}$ (night) for 12 different cities for 3 different days of the week (Tuesdays, Saturdays, and Sundays in the left, central, and right column, respectively) in a given year. Four cities are located in one of the three different latitudinal bands centred at: $37^\circ$N (top); $40^\circ$N (middle); and  $42.5^\circ$N (bottom). For cities lying around $37^\circ$N, the color line associated with their times series are blue, magenta, turquoise and maroon; for cities at $40^\circ$N, the colors are red, orange, brown and yellow; and for cities around $42.5^\circ$N the colors are green, dark green, indigo and gray. Inside each one of the nine plots, the annual behavior of  of $T_{break}$ (lower lines) and $T_{night}$ (upper lines) are shown, for the four different cities located at each band. On every plot, $T_{break}$ and $T_{night}$ show an opposite seasonal variations, with dynamics that appear to counterbalance each other, particularly on Sundays.
From \citet{monsivais2017seasonal}.}
\label{seasonal-pattern}
\end{figure}

Using the dataset \citep{roberts2011costs,saramaki2014persistence} from an 18-month longitudinal study, \cite{aledavood2015daily}  studied circadian patterns in the percentage of calls as a function of the time of the day, and showed persistence of the individual patterns. \cite{cuttone2017sensiblesleep} used the screen on-off data from 400 participants in the CNS~\citep{stopczynski2014measuring} to propose a Bayesian model for estimating individual sleep durations. \cite{monsivais2017seasonal} used data from CDRs \citep{onnela2007structure} of 900,000 users  based in 36 different cities located within the latitudes of 36$^\circ$N and 44$^\circ$N to investigate the influence of seasonality and geography on the sleep and the daily resting patterns in general. They identified inactivity observed in daily mobile phone usage as indication of the human resting patterns. The study showed that the daylight controls human sleeping periods during the nighttime and the seasonal variation of sleep is dependent on the latitude. In the same study they showed that afternoon resting of humans is influenced by temperature, and that a compensatory effect exits between the nightly resting and the resting in the afternoon (see Fig.~\ref{seasonal-pattern}). In another study of the same population, \citet{monsivais2017tracking} investigated the possible age and gender dependencies of sleeping. They also obtained a ``data collapse'' of distributions of the calling activity of cities located at different longitudes, which indicates that the activity of people in urban environments of a country in one time zone follows closely the east-west progression of the sun throughout the year.

We would like to briefly comment on a related area of work that utilizes smartphone applications in sensing of the mental health of individuals. \citet{wang2014studentlife} developed StudentLife, an Android-based application, and conducted a study on 48 student participants to obtain continuous sensing data of them for a period of 10 weeks at the Dartmouth College, UT Austin.  From the sensor data they extracted the behavioural features like mobility, sleep, and type of social conversations (without explicitly analysing the contents), and in parallel ran questionnaires studies. Finally they measured correlations between the sensor data, psychological condition, and academic performance (i.e. grades) of the students. In a similar study, \citet{bogomolov2014daily} tried to predict the reported levels of stress perceived by the participants in terms of their call related features, Bluetooth data, weather conditions and personality traits. There have been studies along similar lines focussing on day-to-day mood~\citep{asselbergs2016mobile}, depression~\citep{saeb2015mobile}, specific mental disorders~\citep{torous2016new}, academic performance~\citep{kassarnig2017academic}, and positioning in the social network~\citep{aledavood2017social}. This is a rapidly developing field\footnote{\url{https://en.wikipedia.org/wiki/MHealth}} in its inception phase with the current focus on infrastructure development~\citep{huguet2016systematic,aledavood2017data}.

\section{Conclusions and future directions}
\label{conclusion}
What we have tried to achieve in this review is the exposition of the approach and methodology of Social Physics to study mobile phone based communication datasets. With this we hope to have demonstrated that Social Physics can give us quite an unprecedented insight into the {\em structure}, {\em function} and {\em behaviour} of social systems and networks of people at different structural and dynamical scales, and this with the help of the data-driven methodologies of {\em data analysis} for making observations or discoveries, {\em modelling} for describing dynamical processes or plausible mechanisms governing the human sociality and their social networks, and {\em simulation} for predicting the behaviour of the social system. 

So far our research quest has concentrated on gaining understanding of human sociality with direct observations and analysis of data from social systems. In order to learn more and get deeper insight into the observed regularities as indication of some kind of governing principles or laws, there is need to reach out and consider whether other complex systems show behavioural similarities, patterns, and universalities rather than differences, variation, and specifics. According to \citet{ball2017complexity} the former viewpoint is akin to physics while the latter is akin to biology. In the studies of complex systems both viewpoints are of course needed as they complement each other and could give us a more solid basis to develop better computational models to be used in simulating or predicting the behavior of social systems more realistically.

The advent and adoption of new channels of communication, is bound to change the distribution of the usage of different communication modalities, especially among younger generations, yet at the same time it widens our scope of human sociality research. Moreover, the approach of leveraging digital information for analysis and modeling of human sociality, embedded in social networks of different scales, is expected to get further developed and refined, for example, by the inclusion of tools like machine learning~\citep{barnett2016feature} which has found a place in the area of condensed matter physics~\citep{carrasquilla2017machine}. Extensive research has already been carried out, focusing in particular on datasets from social media that includes Facebook~\citep{mucha2010community,coviello2014detecting,del2016echo}, Twitter~\citep{golder2011diurnal,beguerisse2014interest,alvarez2015sentiment}, and Wikipedia~\citep{yasseri2012dynamics}. Besides serving as sources of rich datasets, mobile phone networks and other social media platforms have motivated the development of a more generalized framework for the analysis of large-scale socio-technical systems~\citep{holme2012temporal,coviello2014detecting,kivela2014multilayer}.

Human sociality in the contemporary world is no more separable from the existing structures of Internet-driven social technologies. Understanding the structure and dynamics of information exchange in these systems has become pivotal to understanding the functioning of the society, from the perspective of scientists as well as policy makers. There has been serious concern about the rise of adversarial and manipulative behaviour as witnessed for example, in the spread of fake news and the use of so called social bots~\citep{vosoughi2018spread,lazer2018science}. In this respect an interesting avenue of research has opened in the form of human experiments of social phenomena focusing on artificial and online networks, which is in contrast to the idea of passive data collection~\citep{bond201261,centola2015spontaneous,becker2017network,bhattacharya2018role}. The development of hybrid experiments with humans teaming up with bots or algorithms to participate in cooperating tasks, appears to be particularly promising~\citep{shirado2017locally,crandall2018cooperating}. Teams performing efficiently in collaborative tasks are known to have shared mental models. Therefore, the need for modelling prosocial bots has indeed widened the scope of modelling humans.   

\subsection*{Acknowledgements}
KB and KK acknowledge the support from EU HORIZON  2020 FET  Open  RIA  project (IBSEN) No. 662725. KK acknowledges support from Academy of Finland Research project (COSDYN) No. 276439 and from the European Community's H2020 Program under the scheme INFRAIA-1-2014-2015: Research Infrastructures", Grant agreement No. 654024 SoBigData: Social Mining and Big Data Ecosystem" (http://www.sobigdata.eu).


\end{document}